\documentclass{pinchcr}   

\usepackage{graphicx}        
\usepackage{amsmath}

\title{Dynamical Heterogeneities in Grains and Foams}

\author{Olivier Dauchot}
\affiliation{SPEC, CEA-Saclay, 91 191 Gif-sur-Yvette, France}

\author{Douglas J. Durian}
\affiliation{University of Pennsylvania, Department of Physics and Astronomy, Philadelphia, PA 19104-6396, USA}

\author{Martin van Hecke}
\affiliation{Kamerlingh Onnes Laboratory, Universiteit Leiden, Postbus 9504, 2300 RA Leiden, The Netherlands}

\begin{document}

\maketitle

\preface
Dynamical heterogeneities have been introduced in the context of the glass transition of molecular liquids and the lengthscale associated with them has been argued to be at the origin of the observed quasi-universal behaviour of glassy systems. Dense amorphous packings of granular media and foams also exhibit slow dynamics, intermittency and heterogeneities. We review a number of recent experimental studies of these systems, where one has direct access to the relevant space-time dynamics, allowing for direct visualisations of the dynamical heterogeneities.
On one hand these visualisations provide a unique opportunity to access the microscopic mechanisms responsible for the growth of dynamical correlations. On the other hand focussing on the differences in these heterogeneities in microscopically different systems allows to discuss the range of the analogies between molecular thermal glasses and athermal glasses such as granular media and foams. Finally this review is the opportunity to discuss various approaches to actually extract quantitatively the dynamical lengthscale from experimental data.

\begin{figure}[t!]
\centering
\includegraphics[width=.9\textwidth]{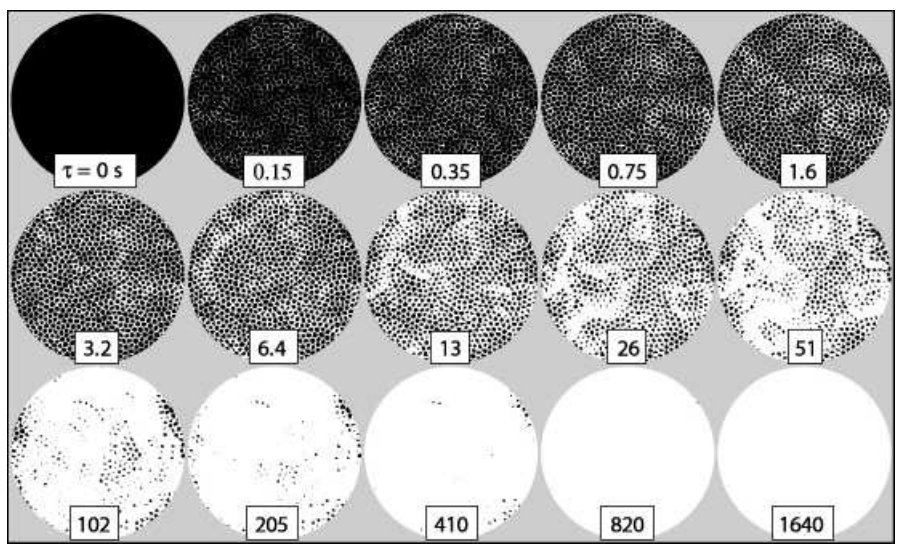}
\caption{An illustration of dynamical heterogeneities in an air-fluidized monolayer of beads at an area
packing fraction of 0.79 (from Ref.~\protect\shortcite{AdamPRE07}). The Voronoi tessellation is computed at each time step and the white regions represent for each bead the space that has come outside of the initial Voronoi cell after a delay $\tau$ as labelled in the figure.}
\label{PA_demo}
\end{figure}

\section{Introduction}
\label{sec:intro}

Granular media and foams exhibit a wide range of complex flow
phenomena, some familiar, some surprising, but often poorly
understood \shortcite{Jaeger1996,CatesPRL98,LiuNagelBook,Duran,midi2004dense,Aranson2006,AndyARFM88,Wilson89,PrudKhan,WeaireBook99,dauchot2007gbr}. The constituents of these materials, macroscopic
grains and gas bubbles, are so large that thermal fluctuations do
not cause appreciable agitations, and their interactions are
dissipative. Hence, when unperturbed, these material jam into a
metastable, disordered state, and to create dynamics energy needs
to be supplied, in the form of, e.g, shearing or vibrations.

The reason why these materials feature in a book on glasses is
that their dynamical behavior often is ``glassy'' --- slow
dynamics, intermittency and heterogeneities are key phrases in
describing their behavior. For example, when repeatedly tapping a
loose packing of grains, their density slowly increases, but
instead of exhibiting an exponential relaxation to an asymptotic density, the
relaxation process is logarithmically slow and, moreover, exhibits
memory effects, evidencing aging (see~\shortcite{dauchot2007gbr} for a review).
In addition, when granular media flow, they typically do so inhomogeneously.
Far away from the main flowing region this results in very slow creeping flows,
where fluctuations are relatively large and where the response is
sluggish. Foams exhibit similar phenomenology.

Dynamical heterogeneities are a key characteristic of glassy
dynamics in thermal systems \shortcite{SillescuSHDinSCL,EdigerSHDinSCL,GlotzerJNCS00,Glotzer03,CipellettiRamos05}, so a
natural question to ask is the nature and organization of the
fluctuations in granular media and foams. As we will outline in
this chapter, the macroscopic glassy features of these materials
are indeed accompanied by heterogeneous fluctuations at the
microscopic, i.e., bubble or grain scale. To wet the readers
appetite, in Fig.~\ref{PA_demo} we show a graphic example of
dynamical heterogeneity in topviews of a system of air fluidized
beads. The coloring in this graph represents the persistent area
order parameter, which quantifies how much the area of the voronoi
cell surrounding a grain has changed in a given lag time.
Note that the dynamics appear homogeneous at
early and long times, but are spatially heterogeneous at
intermediate times - most noticeably for the images at 13~s~$\le
\tau \le$~102~s.

We will focus here on examples of heterogeneous dynamics of foams
and granulates. Since crowding plays a crucial role for these
materials in their dense, glassy phase, one should perhaps not be
surprised that grain and bubble motion is inherently
heterogeneous, and that fluctuations become spatially correlated
--- for one grain to move in a dense system, many other grains
have to get out of the way. Open questions include how far one can
push the analogies between molecular and colloidal glasses on the
one hand, and foams and granular media on the other hand, and what
one can learn about the differences in the heterogeneities in
microscopically different systems.

An important experimental advantage of granular media and foams is
that one has direct access to the relevant space-time dynamics,
allowing for direct observations of the heterogeneous behavior.
Moreover, these materials can be brought close to and often
through a jamming transition, in the vicinity of which dynamical
properties can be expected to change dramatically with control
parameters. Finally, grains and foams have different microscopic
interactions, that are quite well understood --- grains are
essentially undeformable and have inelastic and frictional
interactions, whereas foam bubbles are easily deformed and have
viscous interactions. Hence, by comparing their behavior,
robustness of various glass forming / jamming / heterogeneity
scenarios can be probed. Moreover, the more complex interactions
of grains makes that heterogeneities can be probed in two
physically distinct regimes --- a relatively low packing density
regime where grain interactions are dominated by collisions, and a
higher density regime where frictional contacts dominate the
interactions.

The term ``jamming" has evolved to have many meanings.  It was
originally proposed as an umbrella concept, meant to apply equally to
the glass transition in molecular and colloidal liquids as to the
cessation of flow in grains and foams \shortcite{LiuNagelBook}.   Two
different definitions are offered in Ref.~\shortcite{OHernPRE03}.  The
first is that jamming is said to occur when a system develops a yield
stress, and hence has mechanical rigidity.  However as a practical
matter it is not possible to test whether a material truly has a yield
stress, or whether the stress relaxation time is too long to measure.
So alternatively jamming is said to occur when a systems develops a
relaxation time that exceeds a reasonable experimental times scale,
eg. 1000~s.  This is similar in spirit to defining the glass
transition to occur when the viscosity exceeds $10^{13}$~poise, a
large but arbitrary value.  These two definitions are perfectly
consistent when the relaxation time refers to rigidity, in terms of
the decay time of the macroscopic shear stress relaxation modulus.
However it's a different notion, not always well distinguished in the
literature, to define jamming in terms of the time scale for
microscopic reorganization of structural degrees of freedom like the
set of topological nearest neighbors. Here, we often use jamming in a
rather loose sense, referring to dramatic slowing down of the
dynamics and a qualitative change of the behavior from freely
flowing to being stuck.

The outline of this chapter is as follows.  In section
\ref{sec:agit} we discuss heterogeneities in agitated granular
media. These include air fluidized granular system where the grain
interactions are dominated by collisions, the grains are driven
randomly and isotropically, and the system is quasi-two
dimensional (section \ref{sec:fluidizedbed}), and dense 2D
granular systems where grain interactions are frictional (sections
\ref{sec:micromech} and \ref{sec:criticality}). In section
\ref{sec:micromech} the system is driven by slow oscillatory
shear, and real space rearrangements play a role, while in section
\ref{sec:criticality} the system is driven by horizontally shaking
its support plate, and real space rearrangements are
substantially less than one grain diameter during the duration of
the experiment. In section \ref{sec:heteroflows} we describe observations of
heterogeneities in granular flows in inclined plane, rotating drum
and pile-flow geometries --- for these flows, the grain
interactions are a mix between collisions and enduring contacts,
the flow is driven by bulk shear forces, and the system is three
dimensional. Finally, in section \ref{sec:foams} we describe observations of
heterogeneities and large fluctuations in foams. We close this
chapter by a discussion of commonalities and differences between
these systems, and in the appendix discuss various approaches for
calculating the dynamical susceptibility $\chi_4$ which
quantifies the dynamical heterogeneities. The reader who is not familiar
with dynamical susceptibilities and how they relate to four point dynamical
correlators should refer to this appendix and the first chapters of the present book.

\section{Heterogeneities in Agitated Granular Media}
\label{sec:agit}

\subsection{Growing dynamical lengthscale in a monolayer fluidized
bed} \label{sec:fluidizedbed}

Quasi-two dimensional monolayers of shaken, sheared, and fluidized
grains have been important as model systems, both because it's
difficult to adequately probe an opaque 3d medium but more
crucially because their packing density may be controlled and
varied free from gravitational compaction.  For fluidization, the
setup involves a sub-monolayer of beads that typically roll
without slipping upon a fine screen up through which air is
uniformly blown.  The upward drag need not fully offset gravity to
excite motion - rather, the grains are stochastically kicked
within the plane by the shedding of turbulent vortices.  As
reported in Ref.~\shortcite{AdamPRE06} the air-fluidized grains
therefore experience random ballistic motion at short times and
random diffusive motion at long times.  At intermediate times, and
at high enough densities, the grains also exhibit an interval of
subdiffusive motion where they collide multiple times with a
long-lived set of neighbors.  As the packing density is increased,
the duration of this ``caging" increases until all motion ceases
at random close packing.  Concurrently, there is little change in
packing structure, though the pair correlation function does
exhibit a growing first peak and a split second peak.

\begin{figure}[t!]
\centering
\includegraphics[width=.9\textwidth]{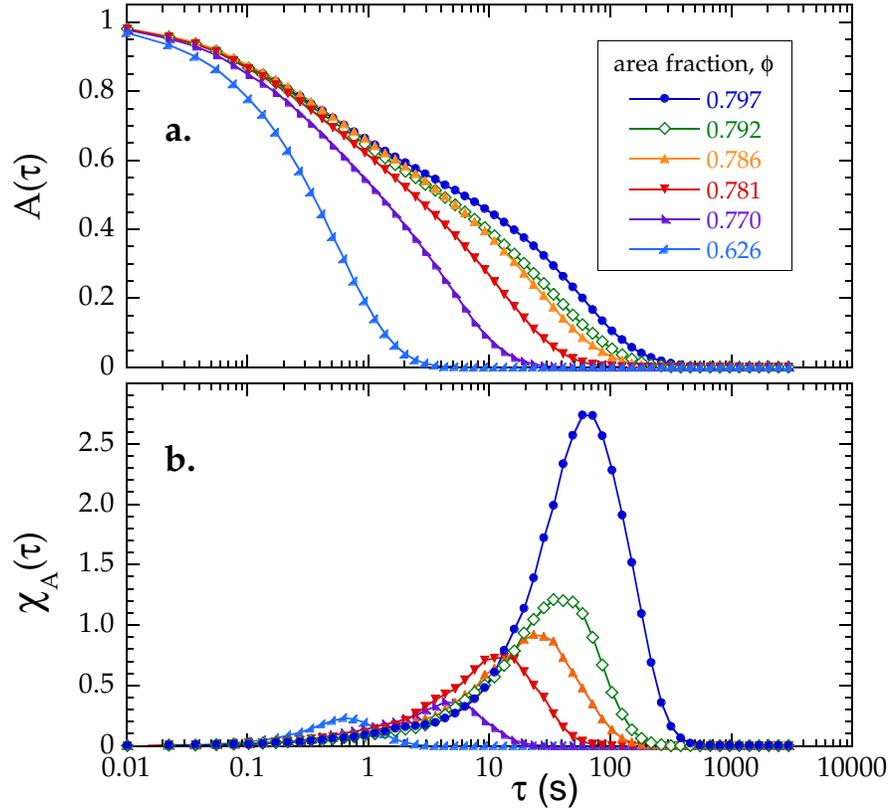}
\caption{(a)Average relaxation, and (b) associated susceptibility, vs delay time for a sequence of packing fractions in the monolayer fluidized bed experiment (from Ref.~\protect\shortcite{AdamPRE07}).}
\label{AX_phi}
\end{figure}

On approach to jamming by addition of grains to increase the
packing density, the dynamics slows down and becomes heterogeneous as illustrated on figure~\ref{PA_demo}.
This has been quantified by measures of the cluster size and of
the chain length for the intermittent fast-moving regions, as well
as through use of dynamic four-point susceptibilities based on
three similar order parameters, which essentially characterize
how much the local structure has evolved (see appendix).
Data for the one based on the overlap of the Voronoi tessalation
(see fig.~\ref{PA_demo}) are plotted in Fig.~\ref{AX_phi} for a sequence
of different packing fractions. The upper plot is a measure of
the temporal relaxation averaged over the whole system. It clearly demonstrates the slowing down of the dynamics when the packing fraction increases. The bottom one shows that the peak in the corresponding susceptibility rises on approach to jamming indicating the presence of a growing length scale~\shortcite{KeysAbateNP07}.

\subsection{Building blocks of Dynamical Heterogeneities}
\label{sec:micromech}

\begin{figure}[t!]
\centering
\includegraphics[width=.45\textwidth]{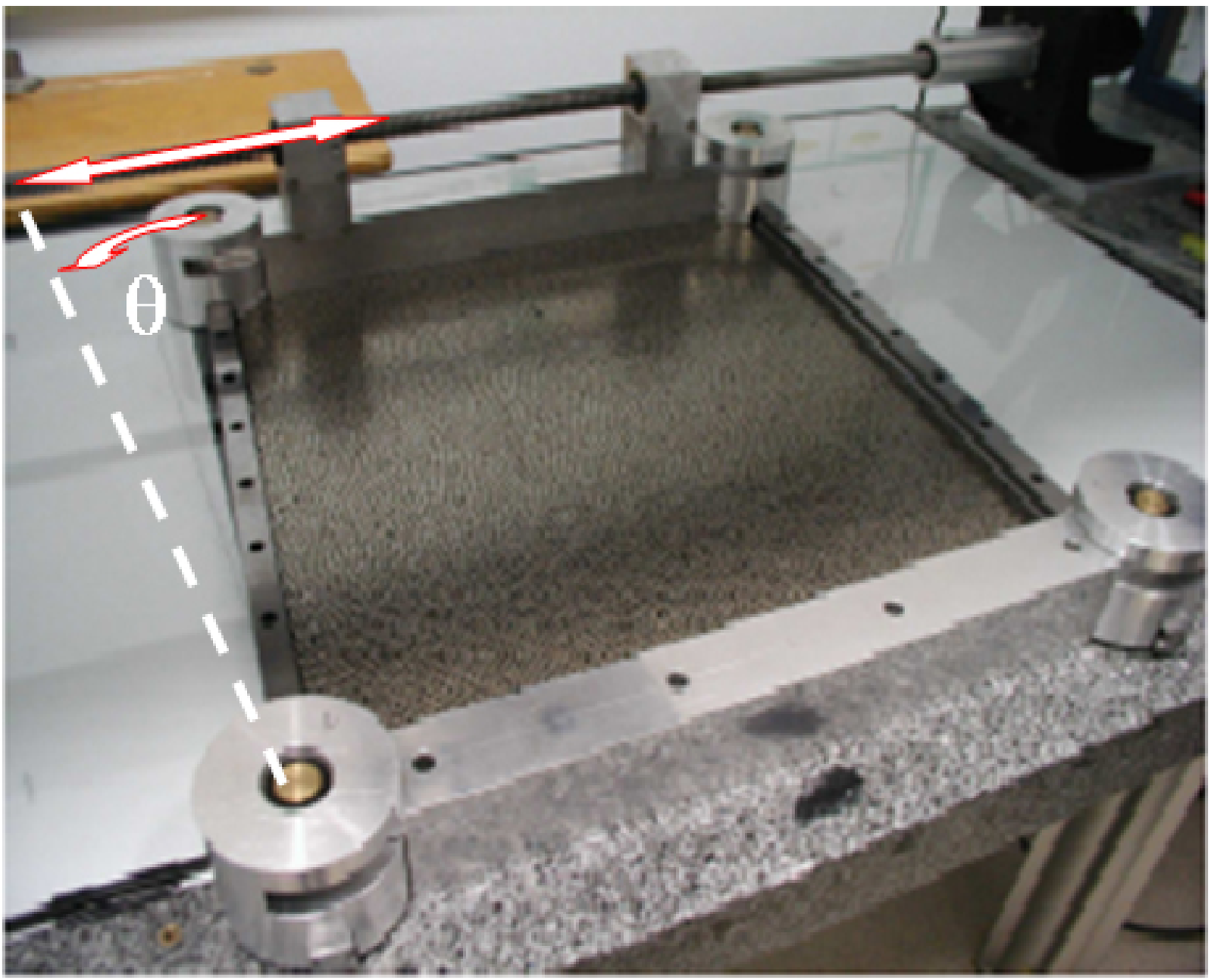}
\includegraphics[width=.45\textwidth]{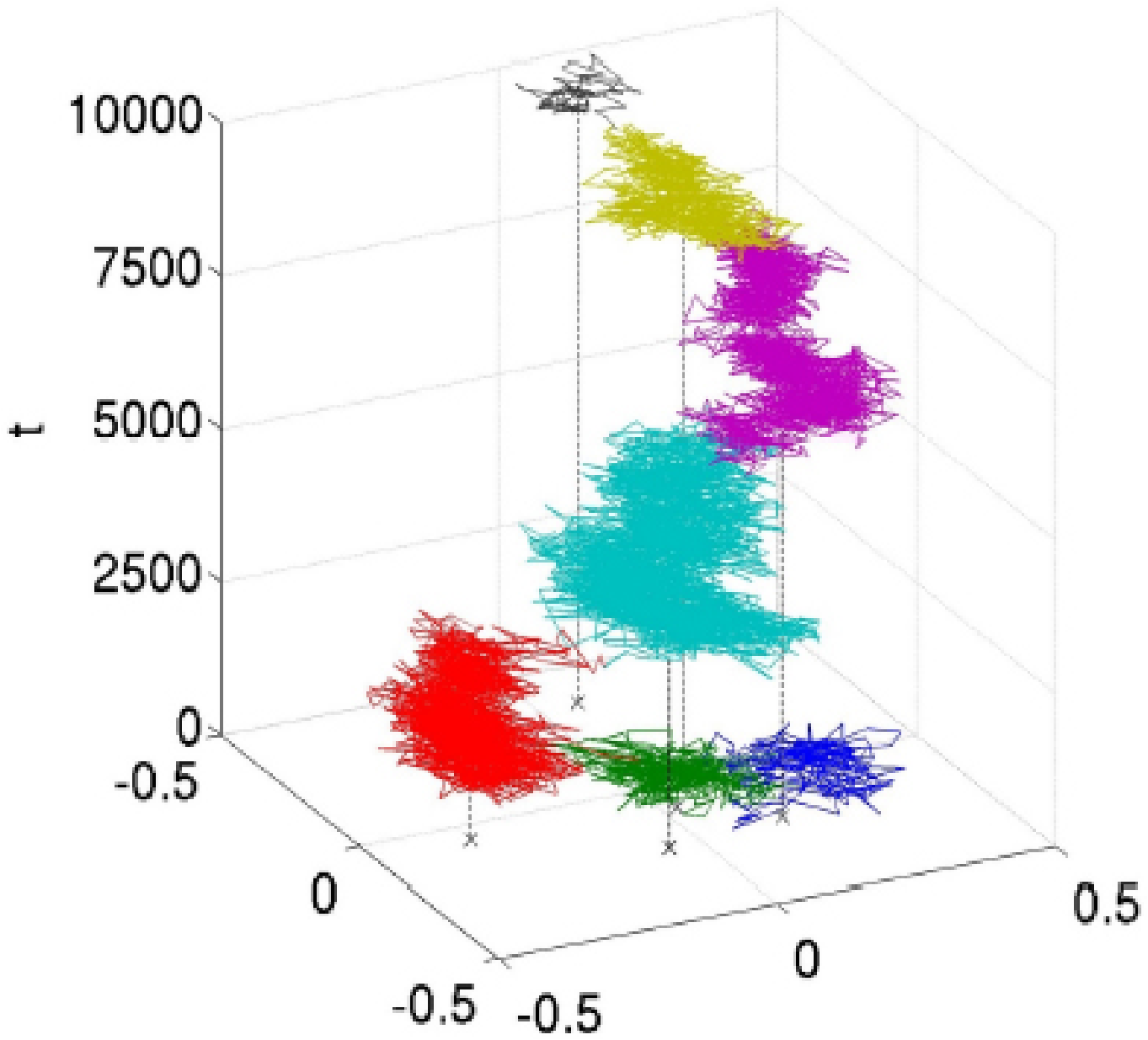}
\caption[]{The cyclic shear cell experiment (from Ref.~\shortcite{marty2005sac}). A bi-dimensional, bi-disperse
granular material, composed of about $8,000$ metallic cylinders of
diameter $5$ and $6$ mm in equal proportions, is sheared
quasi-statically in an horizontal deformable parallelogram. The
shear is periodic, with an amplitude $\theta_{max} = \pm 5^\circ$.
The volume accessible to the grains is maintained constant by
imposing the height of the parallelogram, so that the volume
fraction is a constant $(\phi \simeq 0.84)$. Up to $4000$ grains
located in the center of the device are tracked by a High
Resolution Digital Camera which takes a picture each time the
system is back to its initial position $\theta = 0^\circ$. The
unit of time is then one cycle, one experimental run lasting
10,000 cycles. The unit of length is chosen to be the small
particle diameter $d$.
Images are taken at each cycle and the resulting stroboscopic
trajectories of the grains exhibit typical cages separated by
cages jumps (rhs).}
\label{fig:shearcell-traj}
\end{figure}

Recently, the dynamics of a dense bidisperse monolayer of disks
under cyclic shear has been investigated in Saclay~\shortcite{marty2005sac,dauchot2005dhc,candelier2009bbd}.
The experimental setup is shown in Fig.~\ref{fig:shearcell-traj}-lhs).
The dominant feature of the grain trajectories is the so-called
cage effect (see fig.~\ref{fig:shearcell-traj}-rhs): at short
times, particles are trapped by their neighbors, while at longer times
particles leave their cage and diffuse through the sample by successive cage jumps.
In these experiments the packing fraction is large and cage jumps necessarily
lead to displacements of neighboring particles --- this observation
is at the root of the idea of cooperative motion and dynamical heterogeneities.

\begin{figure}[t!]
\centering
\includegraphics[width=.45\textwidth]{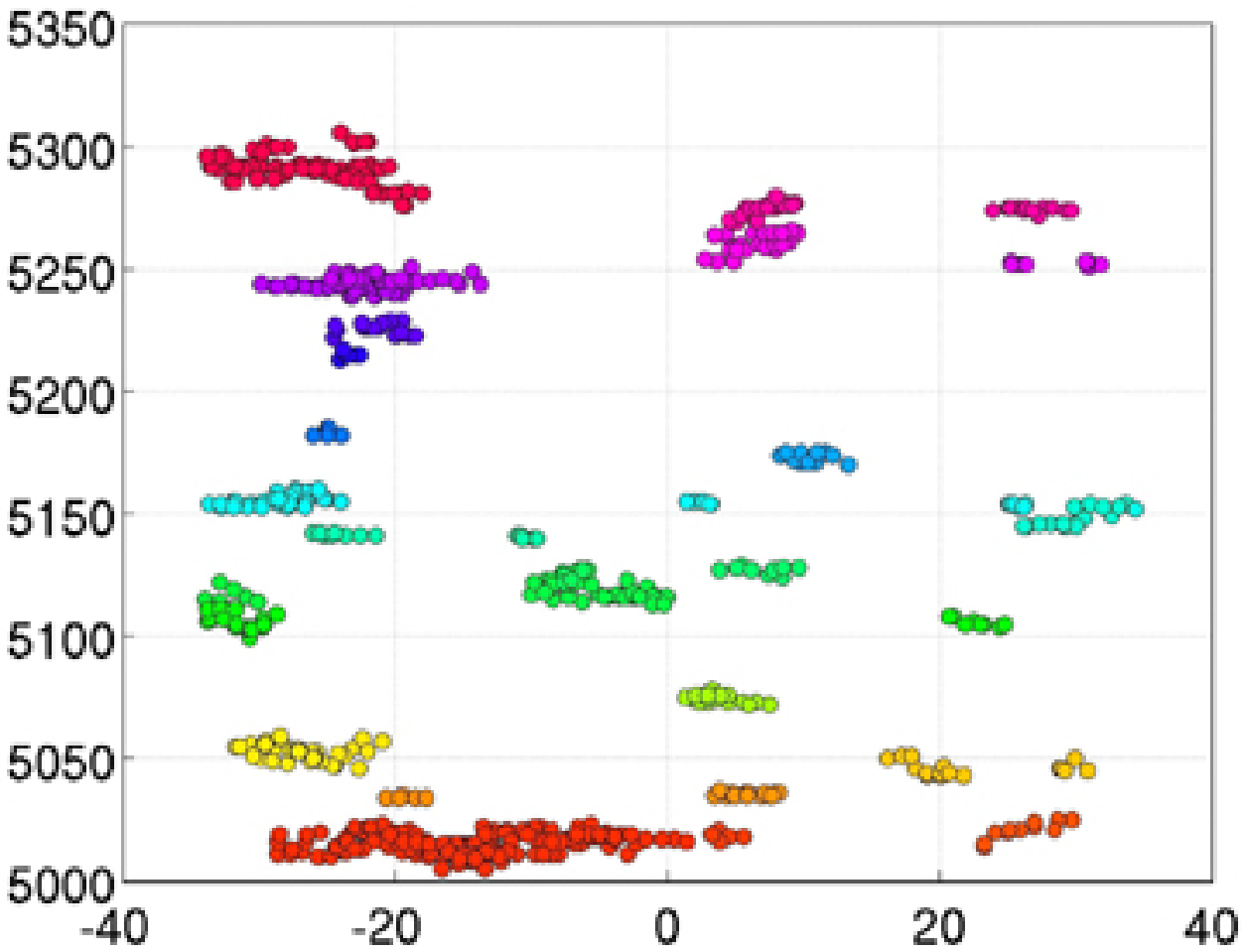}
\includegraphics[width=.45\textwidth]{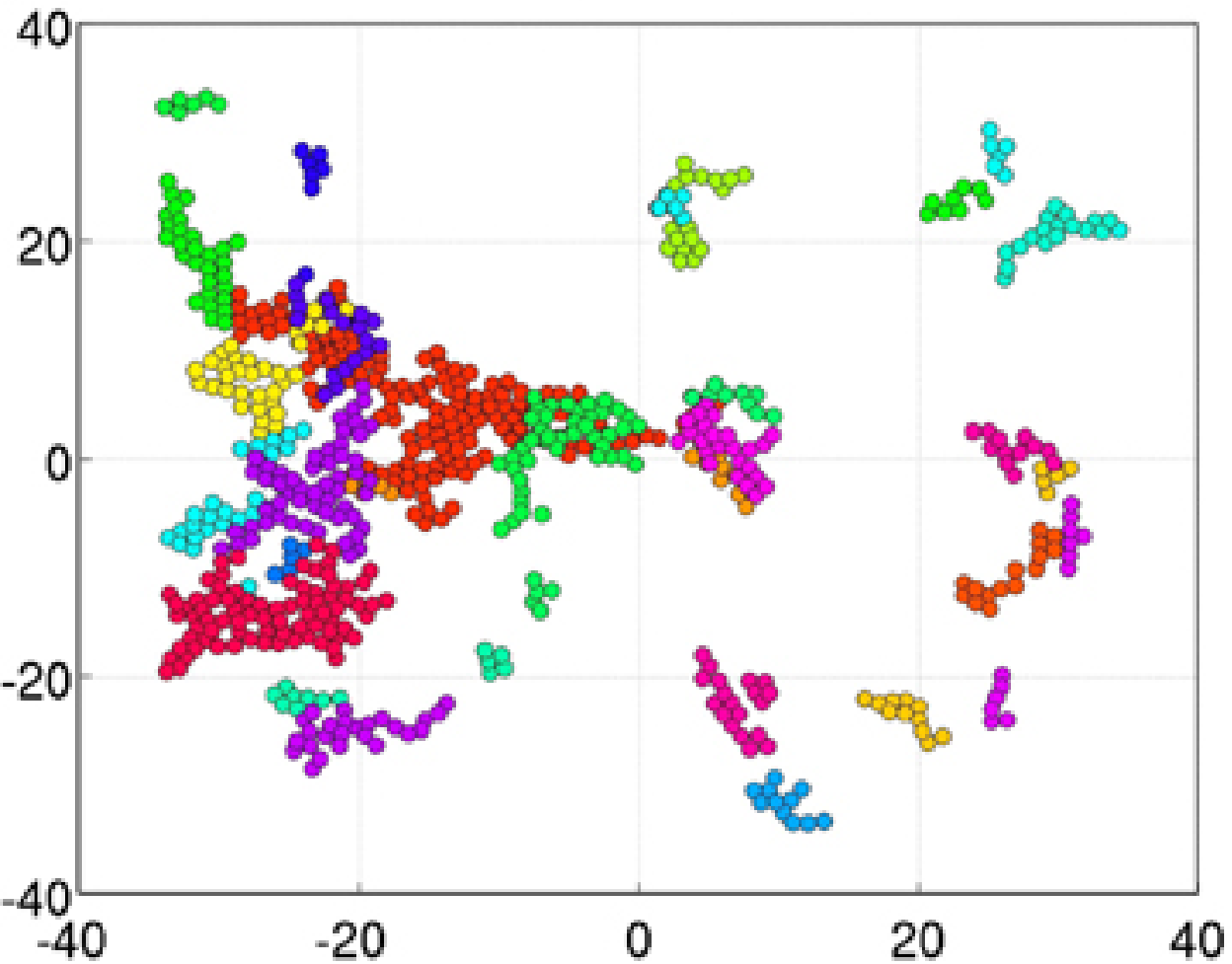}
\caption[]{Spatio-temporal organization of the cage jumps (from Ref.~\shortcite{candelier2009bbd}).
Left: Time of cage jump (vertical axis) vs its x-coordinate (horizontal
axis). Right: Spatial location of cage jumps, showing that cage
jumps facilitate each other to form dynamical heterogeneities.}
\label{fig:cagejumps}
\end{figure}

As is illustrated in figure~\ref{fig:cagejumps}, cage jumps are organized in clusters which avalanche to built
up the long term dynamical heterogeneities ~\shortcite{candelier2009bbd}.
The distribution in space and time of the cage jumps is far from homogeneous.
The left panel of figure~\ref{fig:cagejumps} illustrated that cage jumps form clusters in space,
occurring on a relatively short time scale $\tau_{jump}\simeq 10$.
The cluster size distribution is well described by a power law $\rho(N_c) \approx N_c^{-\alpha}$ where $N_c$ is the number of grains within a cluster and $\alpha \in [3/2-2]$.
The distribution of the lag times separating two adjacent clusters
exhibits an excess of small times as well as an excess of large
times as compared to a Poisonian uncorrelated process, and can be
described by the superposition of two distributions: one for the
long times corresponding to the distribution of the time spent by
the particles in each cage, and one for short delays between
adjacent clusters which suggest a facilitation mechanism among
clusters, the origin of which remains to be found. As a result of
these two timescales, the clusters form avalanches well separated
in time and space.  Finally, selecting a time interval of length
$\tau_{DH}$ corresponding to the timescale for which dynamical
heterogeneities are maximal, initiated at the beginning of a given
avalanche, Fig.~\ref{fig:cagejumps}-rhs displays the spatial
organization of the clusters in the avalanche. One can see how the
clusters spread and built up a region of identical temporal
decorrelation and thereby conclude that the avalanches {\it are}
the dynamical heterogeneities.

Theoretical approaches based on dynamic facilitation usually focus
on kinetically constrained models~\shortcite{ritort2003gdk,garrahan2002prl,toninelli2006jpa}. They
are characterized by a common mechanism leading to slow dynamics:
relaxation is due to mobile facilitating regions that are rare and
move slowly across the system. Here, we find a dynamics
characterized by avalanches inside which clusters are facilitating
each other. However, in the present system facilitation is not conserved
as in kinetically constrained models since the first cluster of an
avalanche is far from any other possible facilitating region.
Recent observations~\shortcite{candelier2009edf} in the fluidized bed experiment described in the
previous section confirm that indeed facilitation becomes less and less
conserved and a less and less significant mechanism when approaching jamming.
Also it has been shown numerically that the mechanisms described in this section also hold in a repulsive supercooled liquid~\shortcite{candelier2009adc}. This is a remarkable fact given the fundamental difference between the athermal granular system and the thermal structural liquid.

\subsection{Criticality across the jamming transition}
\label{sec:criticality}

Once the system has entered the glass phase, its relaxation time
has become much larger than the experimental timescale and it has
fallen off equilibrium. However, one can still increase the
packing fraction under external vibration, up to some value, where
a finite fraction of particles will need to overlap to accommodate
the increase of packing fraction. At that point, the pressure
feels the hardcore repulsion of the grains and jamming occurs.

\begin{figure}[t!]
\centering
\includegraphics[width=.55\textwidth]{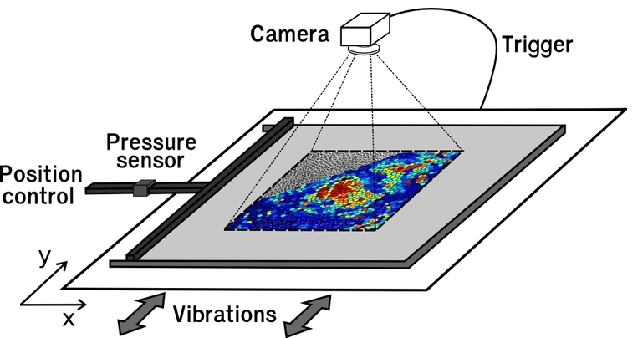}
\includegraphics[width=.35\textwidth]{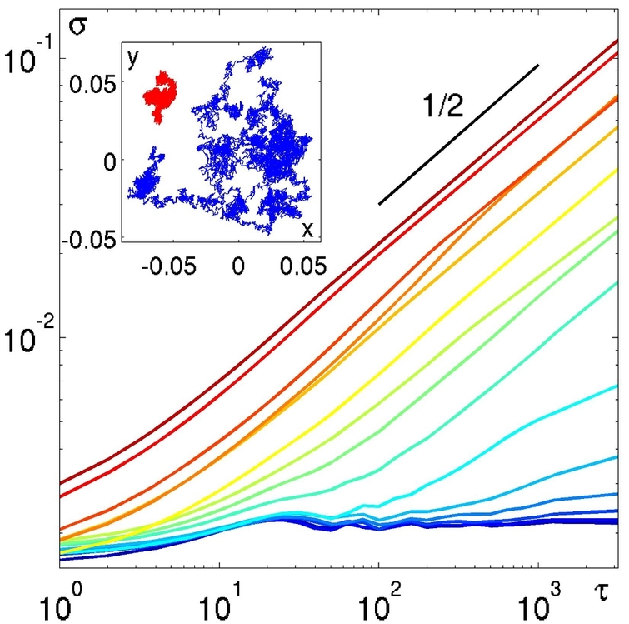}
\caption{A monolayer of bi-disperse grains is driven close to
jamming by successive compression steps under horizontal
vibration (from Ref.~\protect\shortcite{lechenault2008csa}).
Left: Set up. A bidisperse monolayer of 8500 brass cylinders
of diameters $d_{small} = 4\pm0.01 mm$ and
$d_{big} = 5\pm0.01 mm$ laid out on a horizontally vibrated glass
plate $(frequency: 10 {\rm Hz}, amplitude: 10 {\rm mm})$. A lateral
mobile wall allows to vary the packing fraction by tiny amounts
($\delta \phi/\phi \sim 5 \, 10^{-4})$ within an accuracy of
$10^{-4}$. The pressure exerted on this wall is measured by a
force sensor inserted between the wall and the stage. The
stroboscopic motion -- in phase with the oscillating plate -- of a
set of 1500 grains in the center of the sample is tracked by a CCD
camera. Lengths are measured in $d_{small}$ units and time in
cycle units. Right: the root mean square displacements exhibits a
strongly subdiffusive behaviour at short time before recovering
diffusive motion. Note that even for the loosest packing fraction,
the total displacement on the duration of the experiment does not
exceed $0.1$ grain diameter. The inset shows two trajectories,
in blue for the loosest packing fraction, in red for the highest
ones}
\label{fig:vibcell-traj}
\end{figure}

Lechenault {\em et al.} considered the dynamics of a bidisperse
monolayer of disks under horizontal vibration
\shortcite{lechenault2008csa} --- see Fig.~\ref{fig:vibcell-traj}. The
quench protocol produces reproducible, very dense configurations
with structural relaxation time $\tau_\alpha$ much larger than the
experimental time scales. The pressure in the absence of vibration
falls to zero at the jamming transition $\phi_J \in [0.8417,
0.8422]$, and in this system the density can be increased beyond
this transition. One  then observes that long-time correlations,
accompanied by the growth of spatial correlations, are maximal at
$\phi_J$. Here a snapshot of the displacement field reveals the
existence of a super-diffusive motion organized in channel
currents meandering between blobs of blocked particles.

\begin{figure}[t!]
\centering
\includegraphics[width=.9\textwidth]{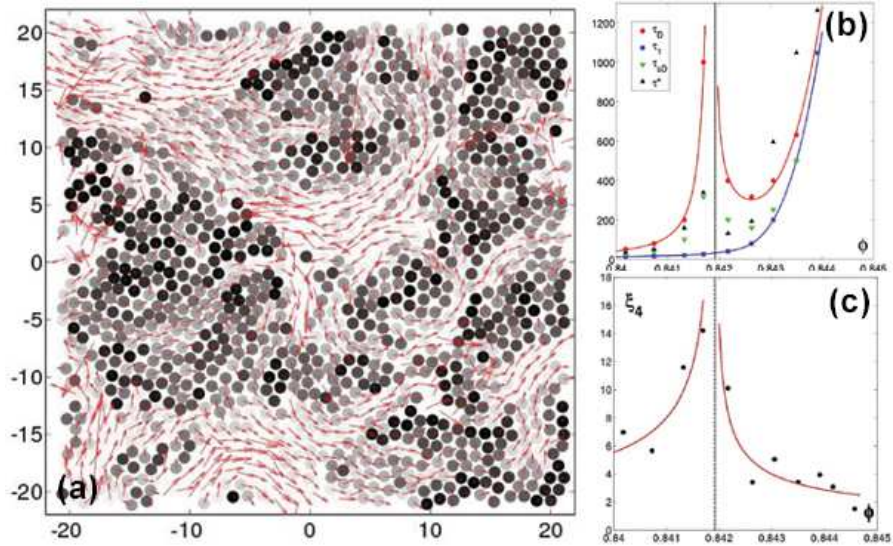}
\caption[]{Criticality at jamming (from Ref.~\protect\shortcite{lechenault2008csa}): displacements on time-scale
$\tau*$ (b) are the most heterogeneous ones; they correspond to
super-diffusive currents of correlated particles (a), which
develop on a length scale $\xi_4$ (c), which strongly increases on
both side of the transition. The time-scale needed to recover
normal diffusion $\tau_D$  (b) exhibits the same sharp peak at the
transition. NB: the displacements have been magnified by a factor
of 50. The color code is black (resp red) for the less mobile
(resp the fastest) particle.} \label{fig:dynhet}
\end{figure}

Figure~\ref{fig:vibcell-traj}-rhs displays the root mean
square displacement as a function of the lag~$\tau$ for various
packing fractions $\phi$. The very small values of
$\sigma_{\phi}(\tau)$ at all timescales are consistent with the
idea that the packing remains in a given structural arrangement.
At low packing fractions $\phi < \phi_J$, and at small $\tau$ the
mean square displacement displays a sub-diffusive behavior before
recovering a diffusive regime at longer timescales. As the packing
fraction is increased, the typical lag at which this cross-over
occurs becomes larger and, at first sight, does not seem to
exhibit any special feature for $\phi\simeq\phi_J$. Above $\phi_J$, an intermediate
plateau appears before diffusion resumes. A closer inspection of
$\sigma^2_{\phi}(\tau)$ reveals an intriguing behavior, that appears
more clearly on the local logarithmic slope $\nu={\partial \log
\sigma_{\phi}(\tau)}/{\partial \log(\tau)}$ (see~\shortcite{lechenault2008csa,lechenault2008lbo}).
For packing fractions close to $\phi_J$ and after the sub-diffusive regime,
the motion becomes {\it super-diffusive} at intermediate times corresponding
to large scale currents shown in figure~\ref{fig:dynhet}-(a).

To characterize the various diffusion regimes, these authors
define three characteristic times: $\tau_1(\phi)$ as the lag at
which $\nu(\tau)$ first reaches $1/2$, corresponding to the start
of the super-diffusive regime, $\tau_{sD}(\phi)$ when $\nu(\tau)$
reaches a maximum $\nu^{*}(\phi)$ (peak of super-diffusive
regime), and $\tau_{D}(\phi)$ beyond which the system recovers the
diffusive regime. These characteristic time scales are plotted as
a function of the packing fraction in Fig.~\ref{fig:dynhet}-(b).
Whereas $\tau_1$ does not exhibit any special features across
$\phi_J$, both $\tau_{sD}$ and $\tau_{D}$ are strongly peaked at
$\phi_J$.

Finally, one can extract the typical size of these
currents by computing the dynamical susceptibility $\chi_4(\tau)$,
which quantifies the number of particles moving in a correlated
manner and exhibits a maximum at $\tau^*$. Interestingly $\tau^*$
behaves like $\tau_{sD}$, a further proof that in the present
case, super-diffusion and dynamical heterogeneities are related.
Recently a deeper analysis of the same data have revealed that the
super-diffusive behaviour must be attributed to the emergence of
Levy-flights in the dispacement distributions rather than to long
time correlations, suggesting the existence of rapid cracks of all
scales rather than the progressive development of soft regions~\shortcite{lechenault2010sdr}.

Another way to characterize the spatial correlation is to compute
the spatial correlator of the displacement field amplitude for a
lag $\tau^*$ (see appendix). The authors could demonstrate that this spatial
correlator also called four-point correlation function
obeys critical scaling $G_4(\vec r,\tau^*) \propto \frac{1}{r^{\alpha}} {\mathcal G}\left(\frac{r}{\xi_4}\right)$, with $\alpha\simeq 0.15$ in the vicinity of $\phi_J$. $\xi_4(\phi)$, plotted on
Fig.\ref{fig:dynhet}-(c), is the length scale over which dynamical
correlations develop. This scaling form, together with the strong
increase of both $\xi_4(\phi)$ and $\tau_{sD}(\phi)$ over a minute
range of $\phi$, is the strongest evidence that the jamming
fraction $\phi_J$ is indeed a critical point, where a static
pressure appears {\it and} long-range dynamical correlations
develop.

\section{Heterogeneities in granular flows}
\label{sec:heteroflows}

\subsection{Flow Rules}

Three different granular flow regimes are to be distinguished.
{\em Rapid flows} are fairly dilute. The main grain interactions
are through collisions, and this regime is described well within
the framework of the kinetic theory
\shortcite{savage1981jfm,goldhirschARFMreview}. {\em Slow flows} are
dense, and the grain interactions are dominated by the frictional
contact forces. This is the regime associated with soil mechanics
\shortcite{nedderman92_book}, although existing descriptions for such
slow flows are rather incomplete and have limited predictive power
\shortcite{fenistein04,deboeuf2005memory}. {\em Liquid-like granular flows} constitute
the intermediate regime, where both inertia and friction are
important and grain interactions are a mix between enduring
contacts and collisions. This last regime has been widely
investigated recently
\shortcite{midi2004dense,savage1998ash,losert2000pds,MLT99,aranson2002ctp,cortet2009rvp,lemaitre2002rds,deboeuf2006flow}.

The crucial progress made recently comes from dimensional analysis
\shortcite{iordanoff2004glt,midi2004dense,dacruz2005rdg} which suggests
that, in simple incompressible unidirectional uniformly sheared
flows, there is only one dimensionless number which governs the
flow: the so-called \textit{Inertial Number}
$I=\dot{\gamma}d/\sqrt{P/\rho}$, which is a function of bead
diameter $d$, grain density $\rho$, global pressure $P$ and global shear rate $\dot\gamma$. The
rheology is then set by requiring that the ratio of shear to
normal stresses is given by an effective friction coefficient
which depends on the inertial number only:
 $\tau/P=\mu(I)$. Such relation has first been
evidenced, numerically, in plane shear
\shortcite{iordanoff2004glt,dacruz2005rdg} and, experimentally, in
inclined plane \shortcite{midi2004dense} configurations. A local
tensorial extension of this relation was recently proposed by Jop
\textit{et al.}~\shortcite{jop2006cld} as a constitutive law for dense
granular flows, and these authors succeeded to fit the surface
velocity profile for a steady unidirectional flow down an inclined
plane with walls. Microscopically, is has been proposed by
Erta{\c{s}} and Halsey~\shortcite{EH02} that the motion of
grains in dense granular flows occurs through clusters, whose size
is controlled by the stress distribution. In the remainder of this
section we will focus now on heterogeneities arising in this flow
regime.


\subsection{Granular Flows in Rotating Drums}

Pouliquen~\shortcite{pouliquen2003fpm} has experimentally studied the
velocity fluctuations of grains flowing down a rough inclined
plane. He has shown that grains at the free surface exhibit
fluctuating motions, which are correlated over a few grain
diameters. Surprisingly, the correlation length is not controlled
by the thickness of the flowing layer but by the inclination only.
The correlation length is maximum at low inclination and decreases
at high inclination, in a similar way as the critical thickness
below which, for a given inclination, the flow stops
\shortcite{DouadyNature99}.

\begin{figure}[t!]
\centering
\includegraphics[width=.9\textwidth]{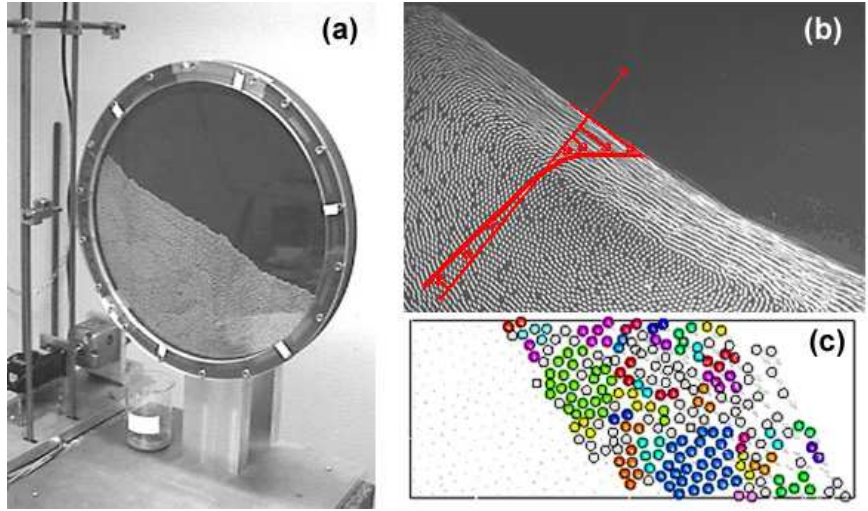}
\caption[]{Rotating drum experiment (from Ref.~\protect\shortcite{bonamy2002msc}). (a) Set up consisting of a
rotating drum of diameter $45$ cm and gap of $7$ mm, half filled
with steel beads of diameter $d = 3\pm 0.05$ mm. A quasi-2D
packing is obtained but with a local 3D microscopic disordered
structure. A fast camera allows to track the 60 percents of the
beads observed through the transparent side wall of the tumbler.
The rotating velocity of the drum is varied from $1$ rpm to $8$
rpm. (b) Linear (resp. exponential) velocity profile in the upper
flowing layer (respectively the lower static layer). (c) Clusters
of beads with correlated velocity fluctuation orientation in quasi
2D flow; typical frame of the clusters for $\Omega = 8 rpm$}
\label{fig:drum-clusters}
\end{figure}

Bonamy et al.\shortcite{bonamy2002msc} have also observed clusters of
particles in the steady flow regime in a rotating drum. In the
recorded region, located at the center of the drum, the granular
surface flow presents the now well known velocity profile, linear
in the flowing surface flow and exponential in the quasi-static
bed (see Fig.~\ref{fig:drum-clusters}-b). By tracking the
particles Bonamy et al. found that the velocity fluctuations of
two beads in contact tend to have correlated orientations.
Figure~\ref{fig:drum-clusters}(c) displays the resulting clusters
--- here clusters are defined as consisting of particles in contact
with velocity fluctuations aligned to within $60^\circ$. The
authors further showed that the distribution of the number of
beads in a cluster is a power law with a cut-off given by the flow
thickness, thereby enforcing an earlier scenario proposed by
Erta{\c{s}} and Halsey~\shortcite{EH02}.

\begin{figure}[t!]
\centering
\includegraphics[width=.9\textwidth]{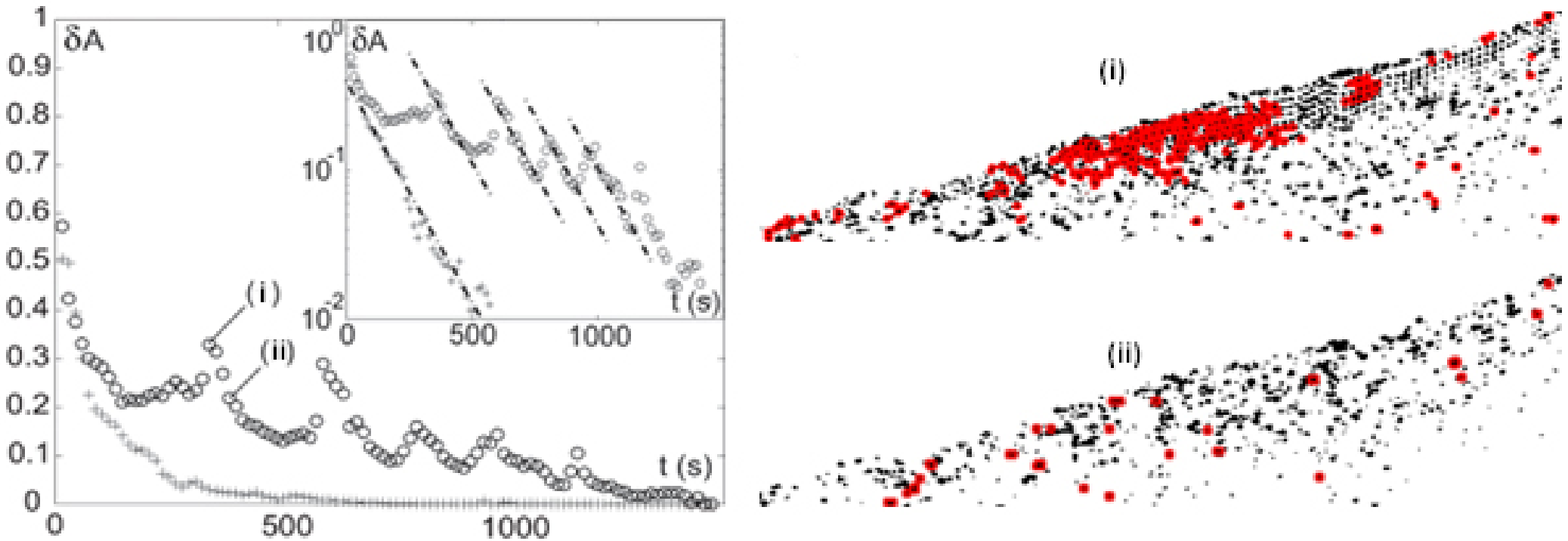}
\caption[]{Intermittent relaxations following a sudden flow
arrest (from Ref.~\protect\shortcite{deboeuf2003jtg}: the evolution of the fraction of mobile grains $ \delta
A(t)$ for two different realizations at similar pile slopes:
$(\bullet) \theta = 15\deg; (+) \theta = 16.5\deg$. Inset is the
log-lin plot of the same data. Notice the exponential decay rate,
which is identical in the monotonous case ($+$) and in the
intermittent case ($\bullet$). On the rhs, displacements in the
pile during a burst event  (i) and during an exponential decay
period (ii). The dark pixels correspond to positions where a
displacement has occurred in the 15 s preceding the considered
time step. The red (light) overlay indicates the pixels, where
displacements have occurred successively during 30 s following the
given time step (see text for details).}
\label{fig:drum-intermittency}
\end{figure}

What are the timescales governing these spatially correlated
clusters? Deboeuf et al.~\shortcite{deboeuf2003jtg} studied the related
question of the typical relaxation times of the granular assembly
inside the drum, once the flow is stopped. For that purpose the
drum is first rotated in the well known regime of intermittent
avalanches, and then stopped just after an avalanche has occurred.
The subsequent relaxation events are then recorded with a standard
CCD camera, which takes images of the pile every 15 s. Denoting
the fraction of beads that have moved between two acquisitions by
$\delta A(t)$, the slow relaxation of the pile can be
characterized.

Two qualitatively different types of behavior can be found, as
illustrated in Fig.~\ref{fig:drum-intermittency}. During the very
first time steps the relaxation process is identical in both
records: the bulk of the pile relaxes rapidly from bottom to top
on time-scales of the order of $15$ s --- typical events
involve isolated bead displacements on short time and length
scales. The relaxation process then slows down in a subsurface
layer of thickness $[10 - 20]$ bead diameters --- this subsurface
layer may relax very differently from one realization to another
(see fig.~\ref{fig:drum-intermittency}).

In one case, one observes a simple exponential decay of the
subsurface layer activity with a characteristic timescale of the
order of $200$ seconds. In the other case, intermittent bursts
interrupt periods of exponential decay, with the same time-scale
as in first case (see inset of
Fig.~\ref{fig:drum-intermittency}-lhs). The competition between
the exponential relaxation and the reactivation bursts results in
a much slower relaxation. A visual inspection reveals that the
reactivation bursts correspond to collective motions of grain
clusters whereas the exponential decay involves individual bead
displacements --- see Fig.~\ref{fig:drum-intermittency}-rhs. For
the case of burst events those displacements persist in time and
are spatially correlated, forming grain clusters.

Altogether the above experiments reveal the existence of
correlated clusters, which seemingly control the thickness of the
steady flow and the relaxation times of the avalanches in the
intermittent regime. Such clusters are purely dynamical in the
sense that they involve spatial correlations of the dynamics, not
of the local structure inside the pile. A characterization of
these dynamical correlations in terms of dynamical
heterogeneities, as introduced for the study of glasses, has not
been done in the case of the rotating drum, at least not in a
systematic manner, but was done for the flow down a pile, as we
will discuss now.

\subsection{Granular Flows down a Pile}

One striking feature of granular flow down a pile is that the flow
near the surface can be very smooth and fluid-like while
simultaneously far below the surface the heap appears to be a
completely static solid. This is true even at very high mass flux,
when the surface flow is steady and independent of time. This
situation has now been extensively studied in a simpler geometry
where the heap is confined in the narrow gap between two
transparent side walls through which the grains may be measured
optically.  Several groups report that the velocity profile along
the sidewall decreases nearly exponentially with depth
\shortcite{PierrePRL00,Khakhar2001,Komatsu2001,Andreotti01,Jop2005,Djaoui05,Richard08}.  Similar localized flow behavior is found
for grains in a rotating drum \shortcite{Rajchenbach1990,duPont2005,cortet2009rvp,midi2004dense},
as well as in Couette \shortcite{Howell99,Meuth00,Bocquet02} and
split-ring \shortcite{fenistein04} cells.  Due to the exponential
character of the velocity profiles for continuous heap flow, the
shear rate is highest near the top free surface where the velocity
is highest, and it decays almost exponentially with depth, too.
Thus the grains experience neighbor changes most frequently and
are most unjammed near the top, and they become progressively
jammed as a function of depth.

The nature of the jamming transition for continuous heap flow,
controlled as a function of depth, was recently studied and
compared with jamming transitions for uniform systems controlled
as a more usual function of temperature or density or shear
\shortcite{Katsuragi09}.  Measurement of the static structure factor
and pair correlation function for grains along the sidewall show
that the spatial arrangement of grains is slightly dilated in the
first layer or two due to saltation.  At greater depths, there is
no noticeable change in structure to accompany the dramatic
decreases in velocity and shear rate.   Such behavior is a
hallmark feature:  glass and jamming transitions are dynamical,
and are not controlled by the growth of a correlation length
associated with instantaneous (static) order.

\begin{figure}[t!]
\centering
\includegraphics[width=0.7\textwidth]{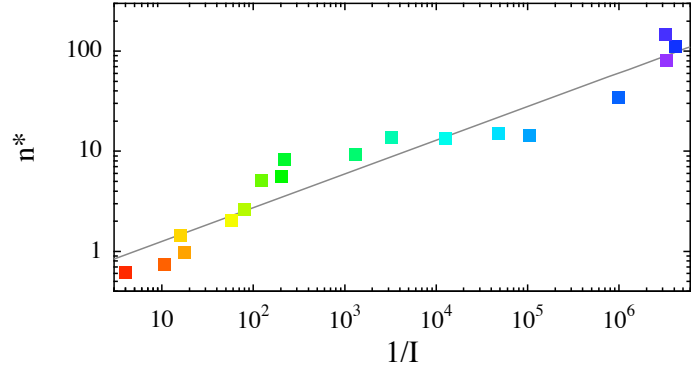}
\caption{Number $n^*$ of grains in a heterogeneity for heap flow
versus inverse inertial number, $I=\dot\gamma d / \sqrt{P/\rho}$,
where $\dot\gamma$ is the shear rate, $d$ is the grain diameter,
$P$ is the local pressure, which depends on depth, and $\rho$ is
the grain density (from Ref.~\protect\shortcite{Katsuragi09}).  The
line is the best fit to a power-law, giving $n^*\sim
(1/I)^{0.33\pm0.02}$ in accord with simulation of a system
undergoing uniform shear \protect\shortcite{Hatano0804}. }
\label{nstar_heap}
\end{figure}

There are two interesting features in the dynamics that both grow
with depth on approach to jamming.   The first is the ratio of the
characteristic grain fluctuations speed, $\delta v$, to the flow
speed, $v_x$; the former is measured by speckle-visibility
spectroscopy, while the latter is measured by particle-image
velocimetry \shortcite{Katsuragi09}.  While both speeds decrease with
depth, the fluctuations do so more slowly in accord with $\delta v
\propto \sqrt{v_x}$.  The relative rise of $\delta v$ over $v_x$
for greater depth signifies an increase in jostling and hence in
dissipation at decrease driving rates, that ultimately results in
jamming; it also means that the flow does not simply slow down
without change in character. In particular, the second interesting
feature is that the character of the dynamics becomes increasingly
heterogeneous on approach to jamming.  This is seen by measurement
of an overlap order parameter and associated susceptibility,
$\chi_4(\tau)$, based on a novel image correlation method that
does not rely on particle tracking \shortcite{Katsuragi09}. At all
depths, $\chi_4(\tau)$ displays a peak vs $\tau$ which is located
very close to grain radius divided by $v_x$, and hence which slows
to longer times at greater depths.  More importantly the height of
the peak, ${\chi_4}^*$, increase nearly exponentially with depth.
This means that the dynamics become increasingly heterogeneous on
approach to jamming.

To compare with other systems, it is more appropriate to consider
the growth in the number $n^*$ of grains in a heterogeneity as a
function of shear rate rather than of depth.  For this, $n^*$ is
computed from ${\chi_4}^*$ as shown in the appendix, and the shear
rate is characterized by the  inertial number \shortcite{midi2004dense}. For
the experiment, $I$ is maximum at about 0.2 near the surface and
decays nearly exponentially with depth.  The scaling displayed in
Fig.~\ref{nstar_heap} of the size of the heterogeneities with
dimensionless shear rate is a power-law relation $n^* \sim
I^{-1/3}$.

%
%

\section{Foams, frictionless soft spheres}
\label{sec:foams}

\begin{figure}[t!]
\centering
\includegraphics[width=.45\textwidth]{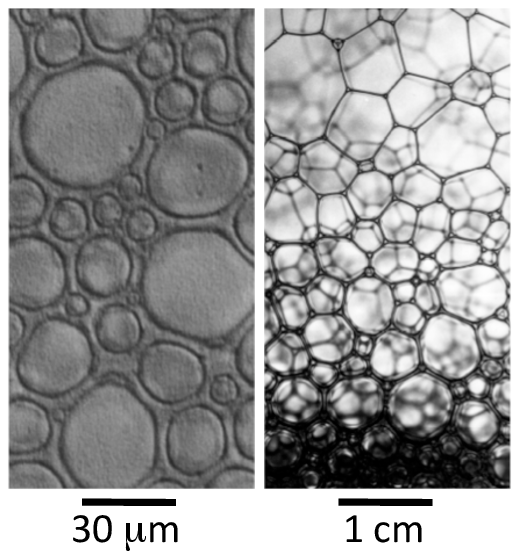}
\includegraphics[width=.45\textwidth]{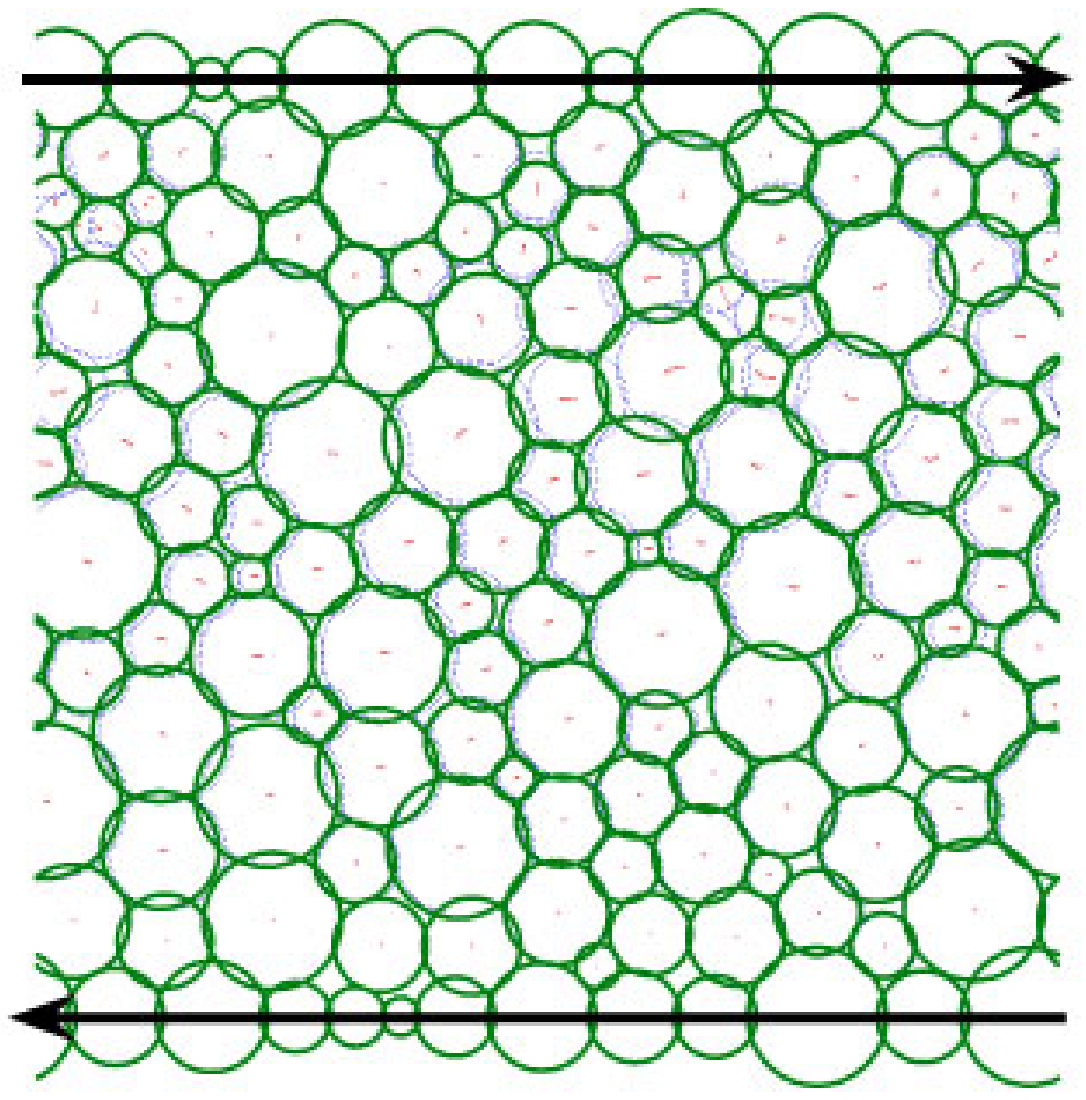}
\caption{Left: Gas bubbles in a shaving cream and in a vial of
soapy water about 30 minutes after shaking.  The former is
about 92\% gas, while the latter has a vertical gradient in
wetness due to gravity. Right: The bubble model as introduced
in ~\protect\shortcite{BubbleModelPRL,BubbleModelPRE} :
bubble positions just before (blue) and after (green)
a shear-induced rearrangement, with trajectory of
the centers shown in red}
\label{bubbles}
\end{figure}

Foams are dispersions of gas bubbles in liquid, stabilized by
surfactants (Fig.~\ref{bubbles}-left) \shortcite{AndyARFM88,Wilson89,PrudKhan,WeaireBook99}.
A crucial parameter is the liquid fraction, or wetness of the
foam, which specifies the volume fraction of the liquid phase.
When the liquid fraction is too large, the individual gas bubbles
do not touch and the material is unjammed
--- one refers to this as a bubbly liquid rather than a foam.
Below a critical liquid fraction --- around 36\% percent for 3d
foams --- bubbles can no longer avoid each other and undergo a
jamming transition. What is particular for foams is that
vanishingly small liquid fractions can easily be reached, where
the foam essentially consists of very thin liquid layers meeting
in quasi 1D plateaux borders, which themselves meet in vertices
--- such foams are called dry foams. In systems under
gravity, drainage can cause gradients in the wetness from very dry
at the top to wet at the bottom.

It is interesting to compare bubbles in foam with grains in a
sandpile. In common, both are comprised of large packing units
which experience negligible thermal motion and which tend to be
jammed. 
But there are many contrasts: First, grains in a pile are
effectively incompressible, and pack at packing fractions {\em
below} random close packing, while bubbles in a foam are readily
deformed and squashed together {\em above} random close packing.
Second, grains are subject to static and sliding friction, as well
as to collisional dissipation, whereas the bubble contacts are
through a liquid film which typically does not support static
friction.

This has important implications for differences between the
jamming of foams and the jamming of grains. Hard grains
essentially are always close to jamming, but due to the friction,
they are not necessarily critical --- the jamming transition for
frictional particles is usually not critical, and is not
characterized by a unique packing fraction or contact number
\shortcite{mvh_jammingreview}. In contrast, the jamming transition for
foams (and emulsions) has all the hallmarks of the theoretically
well-studied jamming of soft frictionless spheres at point J. Here
the jamming transition corresponds to a precise packing fraction
and contact number, and materials near point J exhibit a diverging
length-scale and non-trivial powerlaw scaling of their elastic
moduli \shortcite{ohern2003jzt,mvh_jammingreview}. Some of these
features have, in fact, been discovered first in numerical
simulations of simple models of foams
\shortcite{BoltonPRL90,BubbleModelPRL,BubbleModelPRE} (see fig.~\ref{bubbles}).

Flows of granular media and foams also exhibit essential
differences. Granular flow requires dilatation, and can be
separated in slow and fast flow depending on the role of inertia.
In contrast, foam flows are highly damped due to viscous
interactions, and accomplished by bubble deformation and
rearrangement with no dilatation --- inertia plays essentially no
role.

\subsection{Unjamming of foams}

There are at least three ways to unjam foams. The first is simply
to allow the foam to coarsen: with time gas will diffuse from high
to low pressure bubbles, which generally causes smaller bubbles to
shrink and larger ones to grow.  This is driven by surface tension
through Laplace's law, and serves to reduce the total interfacial
area. As coarsening proceeds, the bubbles rearrange into different
packing configurations and hence can relax macroscopically-imposed
stress.  The time scale for rearrangement can be comparable to
that for size change, as in very dry foams.  But for fairly wet
foams, the rearrangements can be very much faster.  Such
avalanche-like rearrangements are a kind of dynamical
heterogeneity, where a localized region of neighboring bubbles
briefly mobilizes and comes to rest in a new configuration.  For
opaque foams, these events may be captured by diffusing-wave
spectroscopy and its variants \shortcite{DJDsci91,HohlerPRL01,GittingsAO06}. The measured signal gives the time between
successive rearrangements at a scattering site, averaged over both
time and the volume of the sample though which the photons
diffuse.  As the foam coarsens, the time between events is found
to grow as a power-law of time. The DWS signal also includes subtle
contributions from continuous motion of the gas-liquid interfaces,
due to both thermal fluctuations \shortcite{GopalJOSA97}
and also the coarsening process \shortcite{GittingsPRE08,LucaSM10}.

\begin{figure}[t!]
\centering
\includegraphics[width=.9\textwidth]{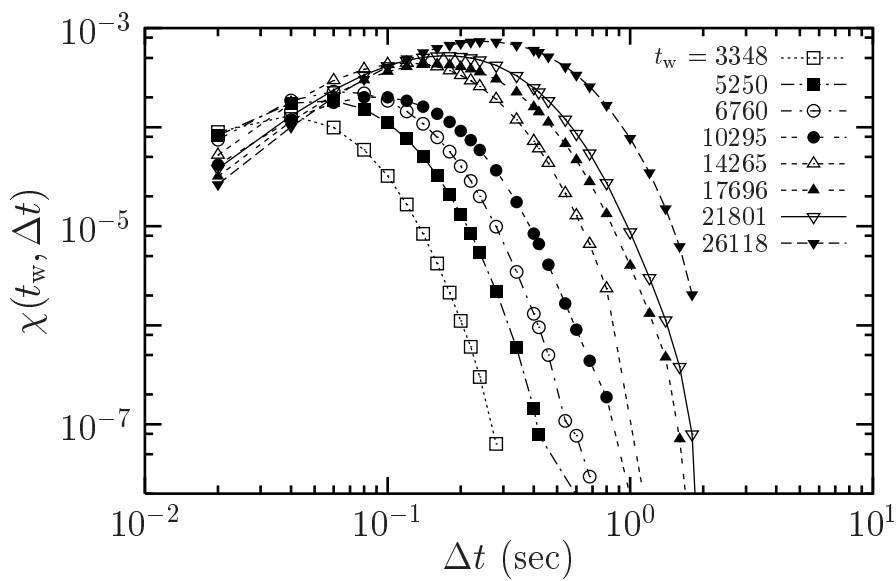}
\caption{Dynamical susceptibility measured by light scattering for a coarsening foam at different wait times $t_w$, labeled in seconds \protect\shortcite{TRConFoam}.  Note that the peaks shifts to longer times and grows in height as the foam ages.
} \label{bubbles2}
\end{figure}

Time-resolved versions of DWS allow further information to be
extracted (see fig.~\ref{bubbles2}).  In Ref.~\shortcite{TRConFoam}, $\chi_4(\tau)$ is measured
by fluctuations in the decay-rate of the DWS correlation function.
As the foam coarsens, the peak location moves to longer times in
accord with the growing time between events.
Furthermore, the peak height also grows -- perhaps because the scattering volume
contains a decreasing number of bubbles and perhaps because the
system is becoming progressively jammed.
In Ref.~\shortcite{GittingsPRE08}, the scattering volume is decreased to
the point that the dynamics of individual rearrangements may be
followed with speckle-visibility spectroscopy.  This allows access
to a second important timescale -- the {\it duration} of the
rearrangement events. In addition, the spatial distribution of successive
events may now be studied with a recently introduced photon correlation
imaging technique \shortcite{LucaPRL09,LucaSM10}.

Foams may also be unjammed by application of shear.  As probed by
DWS  \shortcite{earnshawPRE94,AnthonyPRL95}, the shear-induced
rearrangements appear similar to the coarsening-induced
rearrangements but occur at frequency proportional to the strain
rate.  So coarsening dominates at very low strain rates, small
compared to the reciprocal of the time between coarsening-induced
events.  At very high strain rates, compared to the reciprocal of
the {\it duration} of events, the bubble-scale dynamics are
qualitatively different.  Rearranging bubbles no longer have
enough time to lock into a locally-stable configuration before
having to rearrange again.  Thus successive events merge into
continuous flow, and bubble-bubble interactions are dominated by
dissipative forces rather than surface-tension forces.  This is
evidenced by a change in the functional form of the DWS
correlation function \shortcite{AnthonyJCIS99}, similar to the change
due to diffusive vs ballistic microscopic motion.  The effect of
altered microscopic dynamics on the macroscopic rheology may be
seen to some extent in the shape of the stress vs strain rate flow
curve;  however, it is much more apparent in the transient stress
jump and decay observed when a small step-strain is superposed on
steady shear \shortcite{AnthonyPRL03}.  In particular, the transient
shear modulus and stress relaxation time both decrease vs strain
rate at a characteristic scale set by yield strain divided by
event duration \shortcite{AnthonyPRL03}.

\begin{figure}[t!]
\centering
\includegraphics[width=.8\textwidth]{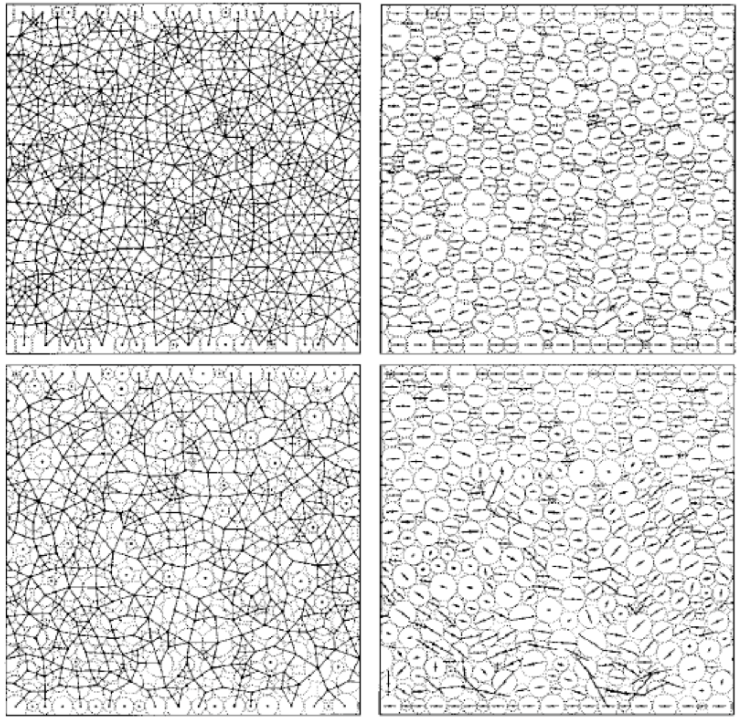}
\caption{Bubble positions and spring network (left column), for
the model of Refs.~\protect\shortcite{BubbleModelPRL,BubbleModelPRE}.
The top row is for a packing fraction of 1 and the bottom row is
for a packing fraction of 0.84.  The right column depicts the
motion that occurs for small-amplitude shear strain, showing that
it becomes more nonaffine on approach to unjamming. }
\label{nonaffine}
\end{figure}

Finally, the third approach to unjam foams is by increasing the
liquid content.  In the dry limit, the bubbles are polyhedral and
separated from their neighbors by thin curved soap films. The
addition of liquid causes inflation not of the films but of the
Plateau borders and vertices at which the films meet. Thus
progressively wetter foams have progressively rounder bubbles,
which unjam when the liquid fraction rises to about 36\% and the
bubbles are randomly close-packed spheres filling about 64\% of
space (in 3d).  This unjamming has been measured in terms of the
vanishing of the shear modulus and the yield strain vs liquid
fraction \shortcite{ArnaudJOR99}.   While rearrangements play no role
in this transition, there is nonetheless interesting changes in
dynamics.  First, as seen in simulation, there is a growing time
scale for stress relaxation \shortcite{BubbleModelPRL}.  This is
accompanied by, and in fact may be due to, bubble displacements
that become increasingly non-affine as the liquid fraction
approaches unjamming \shortcite{BubbleModelPRE} as illustrated on fig.~\ref{nonaffine}.  Non-affine response
has been implicated in a $\sqrt{i\omega}$ contribution to the
complex shear modulus \shortcite{LiuPRL96,AnthonyPRL03}, and in the
non-trivial scaling of the shear modulus with packing fraction
\shortcite{OHernPRE03,WouterEPL09}.

\subsection{Flow of 2D foams}

Experiments on two-dimensional foam under shear have yielded
tremendous insight, in part because the full bubble-packing
structure can be readily imaged and tracked as a function of time
but also because the dry limit may be modeled in terms of
idealized topological features \shortcite{BoltonPRL90,HerdtleAref92,OkuzonoPRE95}.   Pioneering measurements on shear bubble rafts
date back to Argon \shortcite{ArgonMSE79}, who sought analogy with the
flow of metallic glasses.

In recent years, a variety of studies have addressed the flow of
quasi-2D foams, which consist of a single layer of macroscopic $(d
> 1 mm)$ bubbles. Such single layers can be made by freely floating
bubbles on the surface of a surfactant solution \shortcite{wang06}, by
trapping them between a top glass plate and the surfactant
solution \shortcite{katgert08,wang06}, or by trapping
them between two parallel glass plates \shortcite{debregeas}.
The confining glass plates enhance the stability of the foam, but
also introduce additional drag forces that lead to the formation
of shear bands in the foam \shortcite{wang06}.

While the time-averaged flow profiles in such geometries have
received much attention
\shortcite{debregeas,DenninPRL04,wang06,janiaud,katgert08}
here we will briefly outline recent work on the fluctuations
around the average flows. As shown by Debregeas \shortcite{debregeas}
and Lauridsen {\em et al} \shortcite{DenninPRL04}, the instantaneous
flow field exhibits swirly, vortex like motion, commonly observed
in other flowing systems near jamming also. Moreover, T1 events
(local changes in the contact topology) are readily observed in
these systems \shortcite{DenninPRL04,wang06}.

The probability distributions describing the instantaneous bubble
velocities exhibit fat tails  \shortcite{DenninPRE06}. Consistent with
this, M{\"o}bius {\em et al.} established that, for a given local
strain rate, the probability distributions of bubble displacements
exhibit fat tails for short times, develop exponential tails for
intermediate times and finally become Gaussian. The occurrence of
purely exponential distributions at a sharply defined time defines
the relaxation time $t_r$, which coincides with the crossover time
from super diffusive to diffusive behavior, and also with the
Lindeman criterion \shortcite{mobiusunpub}.

Surprisingly,  $t_r$ is not proportional to the inverse of the
strain rate which would be the simplest relation consistent with
dimensional arguments, but instead exhibits a non-linear
relationship with the strain rate. This has a direct consequence
for the probability distributions of bubble displacements taken at
a fixed strain: the width of this distribution grows as
$\dot{\gamma} \rightarrow 0$. This so-called sub-linear scaling,
which has been observed in simulations \shortcite{Ono2003}, implies
that these flows are not quasi-static, but rather that the amount
of fluctuations increases for slower flows --- not dissimilar to
what we discussed above for granular pile flows.

For collections of viscous bubbles with known bubble-bubble
interactions, the balance of work done on the system and the
energy dissipated at the local scale, immediately dictates this
sub-linear scaling \shortcite{Ono2003,Tighe2010}. It has recently been
suggested by Tighe {\em et al} that the nontrivial scaling of the
fluctuations also governs the non trivial rheology of foams ---
where the global relation between strain rate and stress does not
follow directly from the local relation between relative bubble
motion and drag forces \shortcite{olsson,katgert08,Tighe2010}. This
provides an intriguing link between non trivial behavior at micro
and macro scale --- how the spatial organization of the strong
bubble fluctuations associated with sub-linear scaling connects to
dynamical heterogeneities in foams is at present an important open
question.

\section{Discussion}
\label{sec:discussion}

We have discussed the heterogeneities that arise in a variety of
weakly driven systems near jamming. These heterogeneities
have unveiled the existence of a dynamical lengthscale and the associated
timescale responsible for the slow relaxation of these systems.
On one hand the existence of such a lengthscale can be argued to be at the origin
of the quasi-universal behaviour observed in these glassy systems.
On the other hand the observed lengthscale is always rather small,
say smaller than 10 particles diameters and the effect of the
different microscopic mechanisms may still be significant.

For example, for the fluidized grains, interactions are mainly collisional,
for the dense grain systems across the jamming transition their are mainly frictional,
for the pile flows they are a combination of collisions and
enduring contacts, while for foams the interactions are viscous.

Another important difference between these systems is that different
sets of coordinates can be expected to characterize their states.
For example, the structure of collisional grains is set by their
positions only, and structural relaxation will be related to real
space motion and cage breaking, while for dense granular
assemblies that do not show substantial motion of the grains, the
relevant degrees of freedom may be the contact forces, and
relaxations may not be dominated by particle motion but rather by changes
in the contact forces.

The microscopic mechanism of dissipation, which differs between
these systems, may also play an important role. All energy fed
into these systems by shear or agitations needs to be dissipated
by relative motion of neighboring particles. Energy balance then
may lead to the so-called sublinear scaling of fluctuations
\shortcite{Ono2003} --- the details of the energy dissipation then
actually set the width of the distributions characterizing the
fluctuations. Is it also responsible for the spatial organisation
of these fluctuations? More work is needed to clarify this.

As a matter of fact, while it is rather obvious that in dense systems,
local rearrangements will couple to neighboring particles, it is far
from clear what mechanism governs the spatial organization of these relaxation
events. On one hand it is tempting, following recent work~\shortcite{wyart2005ras,brito2006rhs},
to conjecture that the correlated currents observed here are related
to the extended {\it soft modes} that appear when the system loses
or acquires rigidity near jamming. Under the action of a mechanical
drive the system should fail along these soft modes. On the other hand recent
investigations~\shortcite{lechenault2010sdr} suggest that motion of
frictional grains in the vicinity of the jamming transition
can be interpreted as micro-crack events on all scales undermining
the usefulness of harmonic modes as a way to rationalize the dynamics.

Finally, the nature of the relation, if any, between the rheological response
of the materials and the dynamical heterogeneities is far from being understood.
Wether the emergence of a large length scale near jamming controls the rheology
is still a matter of debate. It is also possible that the answer to this question
is different on both side of the transition. Further studies in this matter,
including non-linear rheology and local probe experiments~\shortcite{habdas2004fmp,dollet2005tdf,geng2005sdi,candelier2009cmi}
will certainly contribute to uncover new and probably unexpected effects in this exciting
field of soft matter physics.

\section{Appendix: How to measure $\chi_4$, and the dangers}
\label{sec:Xi4}
The dynamic susceptibility $\chi_4(l,\tau)$ has emerged as a powerful statistical tool for characterizing dynamical heterogeneities \shortcite{SillescuSHDinSCL,EdigerSHDinSCL,GlotzerJNCS00,Glotzer03,CipellettiRamos05}.  However, its definition is somewhat involved and there are pitfalls that must be recognized and avoided if physical meaning is to be extracted from its use.  We offer the following guide to help in this regard.

The first ingredient is an ensemble-averaged dynamical self-overlap order parameter, $Q_t(l,\tau)$, defined such that the contribution from each particle $p$ $Q_{p,t}(l,\tau)$ is some function that decays vs delay time $\tau$ from one to zero as the particle moves a characteristic distance $l$ away from its location at time $t$.
At very short (respectively very large) $\tau$ all particles have moved a distance much less (respectively much larger) than the length $l$ and their contribution to $Q_t(l,\tau)$  is very nearly 1 (respectively nearly 0) with little variance for different start times. By contrast, at intermediate $\tau$ when particles in mobile regions have moved more than $l$ and immobile regions have moved less than $l$, fluctuations in the number of mobile regions cause $Q_t(l,\tau)$ to vary noticeably around its average. In essence, the role of $\chi_4(l,\tau)\equiv N Var(Q_t)$ is thus to capture fluctuations in the number of fast-moving mobile regions. Therefore, as shown in Fig.~\ref{AX_phi}, $\chi_4(l,\tau)$ vs $\tau$ is generally expected to rise from zero to a peak at delay time $\tau*$ when the typical displacement is near $l$ and then decay back to zero.

For the purpose of clarification, let us first introduce a simplified picture of a system of $N$ particles with a fluctuating number $M_t$ of mobile regions of size $n$ and assume that the order parameter is $Q_0$ in all fast regions and $Q_1$ in slow regions. Then the order parameter is given by a weighted average of these values over the total number $nM_t$ of particles in the fast regions and the number $N-nM_t$ of particles in the slow regions:  $Q_t = [f_t Q_0 + (1-f_t) Q_1] $, where $f_t=nM_t/N$ is the fraction of mobile particles. From this one readily obtains the averaged control parameter and $\chi_4$:
\begin{eqnarray}
\bar Q &=& \bar f\, \delta Q + Q_1, \\
\chi_4 &=& N\, Var(Q) = n\, \bar f\, \delta Q^2\, Var(M)/M ,
\end{eqnarray}
where $\delta Q=Q_1-Q_0$ is a measure of how different are fast and slow particles.
If one assumes that there is a large number of mobile regions and that they are decorrelated, then $Var(M)\sim M$ and one can in principle measure the size $n$ of these regions from the above relations.

We now use the above expressions to discuss the precise way of implementing the above procedure.\\

\vspace{3mm}
\noindent {\it Choice of the order parameter:}\\
Most simply, it may be taken as the average over particles of step functions that drops discontinuously from one to zero when a particle moves a distance $l$ \shortcite{GlotzerJNCS00,Glotzer03,KeysAbateNP07}.  A smooth Gaussian cutoff may also be used \shortcite{marty2005sac,dauchot2005dhc}; then the order parameter appears like a dynamic structure factor, which motivates calling $\chi_4(l,\tau)$ a dynamical susceptibility.  The advantage of these two choices is that $l$ may then be adjusted according to the relevant physics.  For example, $l$ approximately equal to particle radius is appropriate for the usual caging where a totally new configuration is obtained when the particles move about a particle size.  Alternative choices have been made in which the cut-off length $l$ is set to probe the topological features of the particle arrangement: the persistent area given by the fraction of space in the same Voronoi cell, and the persistent bond given by the fraction of Voronoi neighbors that remain, after a delay time $\tau$ \shortcite{AdamPRE07}. But for a compressed pack of particles subject to shaking / shearing, a totally new configuration of frictional contacts arises without change in neighbors; then, a much smaller value of $l$ is appropriate \shortcite{lechenault2008csa}. One can also fix $l$ on the basis of the crosscorrelation of grayscale images \shortcite{Katsuragi09}. When in doubt, it is useful to consider the full behavior of $\chi_4(l,\tau)$ vs both $l$ and $\tau$ \shortcite{lechenault2008csa}.

\vspace{3mm}
\noindent {\it Dependence on the packing fraction}\\
As soon as one is interested in the dependence of the size of the mobile regions at the peak of $\chi_4$, $n^*$, on the packing fraction $\phi$ , one should (i) check that the relevant lengthscale $l$ does not vary too much with $\phi$, which is usually the case, (ii) properly normalize $\chi_4$ by $\bar f \delta Q^2$, since the difference in mobility between the fast and the slow particles may well vary with the packing fraction too.

\vspace{3mm}
\noindent {\it Finite size effects}\\
As stated above, it is necessary to have a large enough number of independant mobile regions in order to ensure $var(M)\sim M$. Since one also expects large values of $n^*$ close to the transition of interest, satisfying the above condition requires the use of very large systems, typically of the order of $N=100\, n^*$. These size effects rapidly become critical since the relative error on the measure of $var(Q)$ scales like $\sqrt{N/T}$, where $T$ is the duration of the acquisition.

\vspace{6mm}
One sees that an educated use of $\chi_4$ requires the knowledge of both $\delta_Q$ and $\bar f$, or equivalently the knowledge not only of the ensemble-averaged $Q_t(l,\tau)$, but of all individuals $Q_{p,t}(l,\tau)$. Clearly this is not always the case, and one can already have some insights in the dynamical heterogenities following the above procedure, but keeping in mind the caveats we have just listed. However, when one has the possibility of tracking the particles and thereby has a direct access to the local dynamics, the dynamical heterogeneities are more precisely characterized by calculating directly the spatial correlation of the dynamics, namely the four-point correlator $G_4(r,l,\tau)$:
\begin{equation}
G_4(r,l,\tau)=\left\langle Q_{p,t}(l,\tau)Q_{p',t}(l,\tau)\right\rangle_{d_{p,p'}=r} - \left\langle Q_t(l,\tau) \right\rangle^2 ,
\end{equation}
where $<.>_{d_{p,p'}=r}$ means that the average is computed over all pairs of particles separated by the distance $r$.
The typical lengthscale of the dynamical heterogeneities $\xi_4$ is then readily obtained from the spatial dependance of this correlator. Obviously computing $G_4(r,l,\tau)$ is a more intensive task than the computation of $\chi_4$.

\bibliographystyle{OUPnamed_notitle}
\bibliography{biblio_glass,bibMIDI,bibDJD,bibDJD2,bibmvh,bibglass,fix}
\end{document}